\documentclass[english,aps,prm,twocolumn,amsmath,amssymb,showpacs,superscriptaddress,notitlepage,longbibliography]{revtex4-2}
\usepackage{tikz}
\usepackage{adjustbox}
\usepackage{upgreek}
\usetikzlibrary{arrows.meta,
 chains,
 positioning,
 shapes.geometric
 }
\usepackage{verbatim}\usepackage{tikzscale}
\usepackage{float}
\usepackage{graphics}
\usepackage[version=3]{mhchem}
\usepackage{fvextra}
\usepackage{booktabs,threeparttable}
\usepackage[utf8]{inputenc}
\usetikzlibrary{mindmap}
\usepackage{gensymb}

\renewcommand{\tnote}[1]{\textsuperscript{\TPTtagStyle{#1}}}
\begin{document}

\preprint{1}

\title{Machine Learning-based estimation and explainable artificial intelligence-supported interpretation of the critical temperature from magnetic \textit{ab initio} Heusler alloys data}
\author{Robin Hilgers}
\email[]{robin.hilgers@rwth-aachen.de}
\affiliation{Peter Gr\"unberg Institute and Institute for Advanced Simulation, Forschungszentrum Jülich and JARA, 52425 Jülich, Germany}
\affiliation{Department of Physics, RWTH Aachen University, Aachen, Germany}
\author{Daniel Wortmann}
\affiliation{Peter Gr\"unberg Institute and Institute for Advanced Simulation, Forschungszentrum Jülich and JARA, 52425 Jülich, Germany}
\author{Stefan Bl\"ugel}
\affiliation{Peter Gr\"unberg Institute and Institute for Advanced Simulation, Forschungszentrum Jülich and JARA, 52425 Jülich, Germany}
\affiliation{Department of Physics, RWTH Aachen University, Aachen, Germany}

\date{\today}
\begin{abstract}
Machine Learning (ML) has impacted numerous areas of materials science, most prominently improving molecular simulations, where force fields were trained on previously relaxed structures. One natural next step is to predict material properties beyond structure. In this work, we investigate the applicability and explainability of ML methods in the use case of estimating the critical temperature ($T_\text{c}$) for magnetic Heusler alloys calculated using \textit{ab initio} methods determined materials-specific magnetic interactions and a subsequent Monte Carlo (MC) approach. 
We compare the performance of regression and classification models to predict the range of the $T_\text{c}$ of given compounds without performing the MC calculations. Since the MC calculation requires computational resources in the same order of magnitude as the density-functional theory (DFT) calculation, it would be advantageous to replace either step with a less computationally intensive method such as ML. We discuss the necessity to generate the magnetic \textit{ab initio} results, to make a quantitative prediction of the $T_\text{c}$. We used state-of-the-art explainable artificial intelligence (XAI) methods to extract physical relations and deepen our understanding of patterns learned by our models from the examined data.
\end{abstract}
\maketitle
\section{Introduction}
\label{Intro}
Machine-Learning (ML) modeling has shown to yield promising results in various scientific sectors and applications~\citep{Pharma,Cancer,Mood}. The ability of flexible learning algorithms to recognize patterns, adapt to data properties, and tackle challenges such as regression, classification, and clustering has established an additional scientific paradigm of data-driven science besides the traditional paradigms of experiments, theories, and simulations. Data-driven science essentially shifts scientific problem-solution strategies for predictions from problem-specific models to versatile data-based models~\cite{DatDriv,MLRoadMap,LevelsOfDatDriv}. This is also the case for a plurality of materials science applications ranging from superconductivity~\cite{SupercondML}, molecular dynamics~\cite{Deringer2018}, materials synthesis, and design~\cite{MatSynthesisML}, knowledge discovery through data mining~\cite{DataMiningMatScience}, entropy changes~\cite{MLEntropy}, and other topics for both properties and materials prediction~\cite{MLReview,RecentMLRev,MLRoadMap}. For some of the mentioned applications, \textit{e.g.}\ in some molecular dynamics simulation applications~\cite{Deringer2018}, lightweight and computationally inexpensive ML-based approaches were able to virtually replace established techniques, while in other applications ML-based approaches complement existing methodologies~\cite{MLRoadMap}. Data mining-related techniques have shown to be powerful tools in the hands of scientists to discover relations within data, even in the materials science community~\cite{DataMiningMatScience}.

There are a multitude of magnetic properties to investigate, many of which are traditionally described by complex models based in part on the quantum mechanics of the many-electron problem. Within the set of magnetic properties, the critical temperature, also known as the Curie temperature in the context of ferromagnetic materials, represents a key characteristic in both fundamental physics and practical applications. It provides valuable insights into the transitions between different magnetic phases and guides the design and optimization of magnetic materials for technological use. For example, in the design of magnetic materials for the energy use sector of the economy~\cite{Gutfleisch:11}, \textit{e.g.}\ electric power generation, conditioning, conversion, transportation, or the information sector of the economy, \textit{e.g.}\ spintronics~\cite{PhysRevB.98.224407} or magnetic storage devices (like hard drives), the critical temperature determines the maximum operating temperature where magnetic data storage remains stable. Typical application demands necessitate critical temperature values significantly exceeding room temperature~\cite{400K}. Hence, in order to conduct application-oriented material screening studies at a high-throughput scale for materials discovery, a lightweight method is required to predict whether the critical temperature of a compound meets the requirements set by the applications. Existing works, mostly focused on the Curie temperature in ferromagnetic materials~\cite{SanVitSimilar,SanVitoML}, while the more general concept describing a wide range of magnetic phases, including ferromagnetic, anti-ferromagnetic, ferrimagnetic, and spin-spiral type ordering is the critical temperature of the phase change transition of the ordered magnetic to a non-magnetic state represents the field of interest in this study.

Within the phase space of magnetic materials, the Heusler (and Heusler-like alloys) alloys are known to represent candidate materials for various technical applications, as the material class of Heusler \cite{HeusOrig,HeusOrig2} alloys (as \textit{e.g.} the ordered $\mathrm{L2}_1$ phase) and related disordered phases (such as \textit{e.g.} A2 and B2 phases) are known to exhibit many interesting properties including superconductivity~\cite{SuperC}, piezoelectricity~\cite{Roy2012}, rare-earth free permanent magnets~\cite{MagnetsHeus}, and half-metallicity~\cite{HMHeus}. The combination of multiple properties in a single compound such as \textit{e.g.}\ both half-metallicity and magnetic stability allow for the occurrence of spin-polarized charge currents, which are a topic that is actively investigated by the scientific community for applications in spintronics~\cite{SpinCurrent,HeusSpint}. By including not only the ordered but also disordered phases and quaternary Heusler alloys, the phase space of possible compounds increases drastically in comparison to existing works like \textit{e.g.}\ \cite{SanVitoSmallTheorySet}, which restricts the phase space to pure transition-metal Heusler alloys. However, as a Heusler alloy's structure is defined by the individual compound's lattice site constituents, the lattice constant, and the symmetry group alone, the structural parameters that have to be considered by a model in order to describe such a system are very limited. 

\begin{figure}[ht]
\resizebox{\columnwidth}{!}{\begin{tikzpicture}[
    node distance = 5mm and 7mm,
      start chain = going right,
 disc/.style = {shape=cylinder, draw, shape aspect=0.3,
                shape border rotate=90,
                text width=16mm, align=center, font=\linespread{0.8}\selectfont},
  mdl/.style = {shape=ellipse, aspect=2.2, draw},
  alg/.style = {draw, align=center, font=\linespread{0.8}\selectfont},     arrow/.style = {thick, -Stealth},
                    ]
\node (n1) [disc] {Magnetic \\Heusler\\ Structures };
\node (n21) [alg, above  right = 2.5mm and 8mm of n1]  {Ab-Initio KKR\\ Calculation};
\node (n22) [alg, right =8mm of n1]  {Ab-Initio KKR \\ Calculation};
\node (n32) [alg, right =8mm of n22]  {Machine-Learning \\ Model};
\node (n3) [alg, right = 12mm of n21]  {Monte-Carlo\\ Simulation};
\node (n4) [disc, right = 8mm of n32] {$T_c$};
\node (n23) [alg, below right = 2.5mm and 22mm of n1]  {Machine-Learning Model};
\draw[arrow] (n1.north)  -- (n21.west);
\draw[arrow] (n21.east)  -- (n3.west);
\draw[arrow] (n1.east)  --( n22.west);
\draw[arrow] (n1.south) -- (n23.west);
\draw[arrow] (n22.east) -- (n32.west);
\draw[arrow] (n32.east) -- (n4.west);
\draw[arrow] (n3.east) -- (n4.north);
\draw[arrow] (n23.east) -- (n4.south);
    \end{tikzpicture}}
\caption{Schematic depiction of the layered $T_\text{c}$ determination with different ML integration levels.}
\label{SchemeDFT}
\end{figure}
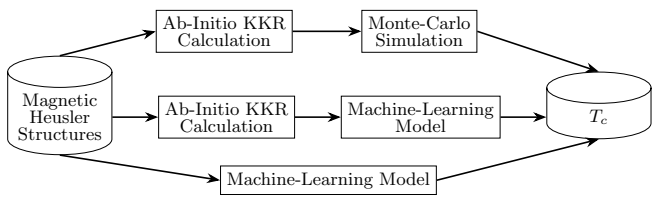

In this paper, we aim to demonstrate the advantages offered by ML, replacing traditional $T_\text{c}$ determination using density-functional theory (DFT) and Monte Carlo (MC) simulations. We focus on the prediction of the magnetic critical temperature for ordered (Including the phases $\mathrm{L2}_1$, $\mathrm{C1}_b$, Y, and XA) as well as disordered (Including the phases A2 and B2) magnetic Heusler alloys. The critical temperatures were determined in a two-step process of an \textit{ab initio} KKR-GF~\cite{KKR} DFT simulation followed by an MC simulation of the $T_\text{c}$ as depicted in the top path of Fig.~\ref{SchemeDFT}.
As both steps are comparable in computational cost, we apply our modeling for the whole process as well as only the MC step, taking advantage of magnetic results obtained in the \textit{ab initio} step. 

Beyond that, we discuss the impact of magnetic features for the prediction of high $T_\text{c}$ materials and the usability in high-throughput materials screening applications, which do not include DFT-originated features in the first place. This discussion is heavily assisted by the use of explainable artificial intelligence (XAI) techniques, which we demonstrate to be able to explain model predictions based on materials science data and visualize relations in the training data captured by the ML model~\cite{XAIMatSc}.

\section{Methods and Materials}
\label{Meth}
\subsection{Data Processing \& Cleaning}
The examined data was collected at our institute and published as the Jülich-Heusler-magnetic-database (JuHemd)~\cite{DataHeusler2022}. It provides not only 
structural and stoichiometric information on the Heusler compounds, but also magnetic data obtained by DFT and Monte Carlo simulations. The target quantity we want to predict in our modeling is the critical temperature $T_\text{c}$ of the magnetic ordering. While the JuHemd contains experimental values as well as 
those based on DFT simulations 
using GGA~\cite{GGA} and LDA~\cite{LDA} exchange-correlation functionals, we restrict our analysis to the GGA-based values as these are provided for most compounds and provide the most homogeneous data quality.

As a first preparation step, we extract the $T_\text{c}$ values together with a set of descriptors for each compound in the database. All information was encoded into a numerical representation and made available for the modeling process. Using the provided metadata to augment the information with additional atomic features, we finally obtain a set of 118 descriptors, as listed in table \ref{Tab}. Before any ML modeling is performed, these descriptors $\{x_i\}$ are then transformed to a standardized form \begin{equation}
    \{z_i\}= \frac{\{x_i\}-\mu_i}{\sigma_i}
\end{equation} 
using the mean $\mu_i$ and standard deviation $\sigma_i$ of the $i$-th descriptor in the training set. 
\vspace{1cm}

\begin{table*}
\centering
          \begin{threeparttable}[b]
            \caption{List of all features which are contained in the processed data and their corresponding explanation. \\For all features that were directly derived from the JuHemd,\\ the JuHemd label has been used. Also, JuHemd labels have been included which were used\\ to construct processed quantities even though the original label is not\\ included in the processed data set due to the format, the quantity is given in the JuHemd.}\label{Tab}
            \begin{tabular}{ll}
              \toprule
              Label & Description \\
              \midrule
              \Verb|lattice_constant|\tnote{*} & Lattice constant of the Heusler \\
               \Verb|resval|\tnote{*}& $T_\text{c}$  \\
              \Verb|etotal (Ry)| \tnote{*}& Total energy of the compound $E_{Tot}$ \\
              \Verb|formula|\tnote{*} & Chemical formula of the compound  \\
              \Verb|Ferro Density|\tnote{\textdagger} & Fraction of ferromagnetic elements (\ce{Fe}, \ce{Ni}, \ce{Co}) in the Compound \\
              \Verb|Rare earth Materials Density|\tnote{\textdagger} &  Fraction of rare earth components in the Compound\\
             \Verb|Symmetry Code|\tnote{\textdagger} & An integer encoding the Heuslers symmetry group \\       \Verb|Individual Magnetic Moments|\tnote{\textdagger} & Individual magnetic moments $m_i$ of all constituent atoms  \\
                            \Verb|Absolute Magnetic Moments| \tnote{\textdagger} & Individual absolute magnetic moments $|m_i|$ of all constituent atoms  \\
               \Verb|Total magnetic moment|\tnote{\textdagger} & $M=\sum\limits_i m_i$ \\
               \Verb|Sum of absolute magnetic moments|\tnote{\textdagger} & $M_{Abs}=\sum\limits_i |m_i|$ \\

               \Verb|Magnetic State|\tnote{\textdagger} & Integer encoding the magnetic state (Ferro, AFM, and Spin-Spiral) \\
               \Verb|Stochiometry|\tnote{\textdagger} & 5-Digit integer encoding the stochiometry of the compound \\
               \Verb|Density by Atomic Number|\tnote{\textdagger} \tnote{1} & Fractional density of each atomic number is encoded by an individual descriptor  \\
               \Verb|Atomic Number|\tnote{\textdaggerdbl} &  Atomic number of the constituents $Z_i$\\
               \Verb|Number of Neutrons|\tnote{\textdaggerdbl} & Number of neutrons of the constituents\\
                \Verb|Nominal Mass|\tnote{\textdaggerdbl} & Nominal mass of the constituents atoms\\
                 \Verb|Number of Electrons|\tnote{\textdaggerdbl} &  Number of electrons of the constituents\\
                  \Verb|Exact Mass|\tnote{\textdaggerdbl} & Exact mass of the constituents atoms \\
                   \Verb|Atomic Radius|\tnote{\textdaggerdbl} & Atomic radii of the constituents atoms\\
                  
                   \Verb|Number of Valence Electrons|\tnote{\textdaggerdbl} & Number of valence electrons of the constituents atoms $e^{val}$\\
                   \Verb|Covalence Radius|\tnote{\textdaggerdbl} & Covalence radius of the constituents atoms\\
                   \Verb|Period|\tnote{\textdaggerdbl} & Period number in the PSE of the constituents atoms \\
                   \Verb|Electronegativity|\tnote{\textdaggerdbl} & Electronegativity of the constituents atoms $\chi_{i}$\\
                   \Verb|Van der Waals Radius|\tnote{\textdaggerdbl} & Van der Waals radius  of the constituents atoms $r^{vdw}_i$\\
                   \Verb|Electron Affinity|\tnote{\textdaggerdbl} & Electron affinity  of the constituents atoms $E_{ea\ i}$\\
              \bottomrule
            \end{tabular}
            \begin{tablenotes}
            \item[*] Available directly from JuHemd
            \item[\textdagger] Constructed descriptors
            \item[\textdaggerdbl] Added atomic descriptors - most have four entries per compound
            \item[1] This feature has as many entries (columns) as the JuHemd contains a plurality of unique elements from the PSE
            \end{tablenotes}
          \end{threeparttable}
\end{table*}
\vspace{1cm}

Only those compound entries have been included which contain all of the above-mentioned entry labels. Incomplete data points have not 
been used. Additionally, only magnetic alloys are selected. We chose the magnetic cutoff 
to be \begin{equation}
    \sum\limits _{i}|m_i|>0.1 \mu_B
\end{equation} where the $m_i$ denotes the magnetic moment of the atom on site $i$ in the compound's molecular formula. Similarly, we did not include compounds with a simulated $T_\text{c}=0\ \mathrm{K}$. This leaves us with a final data set size of 408 Heusler compounds. 

Since, during the data processing, 
incomplete data points for Heusler compounds are removed, there are some elements from the periodic table that are contained in the original JuHemd but are not contained anymore in the processed data. The corresponding densities of these atomic numbers, which originate from these removed elements, represent descriptors with zero variance in every compound. Such descriptors are removed before further processing, as they are 
meaningless for the ML training and evaluation process. In this paper, of the 118 descriptors, there are 11 descriptors in the data set with zero variance, which are hence removed.
The whole data order has been randomized in order to avoid the clustering of similar data points due to the alphabetical order. This enforces 
homogeneity of the data set, which is necessary for the Cross-Validation (CV)~\cite{EleStat} model evaluation to be meaningful. 

\begin{figure*}[]
\includegraphics[width=\textwidth]{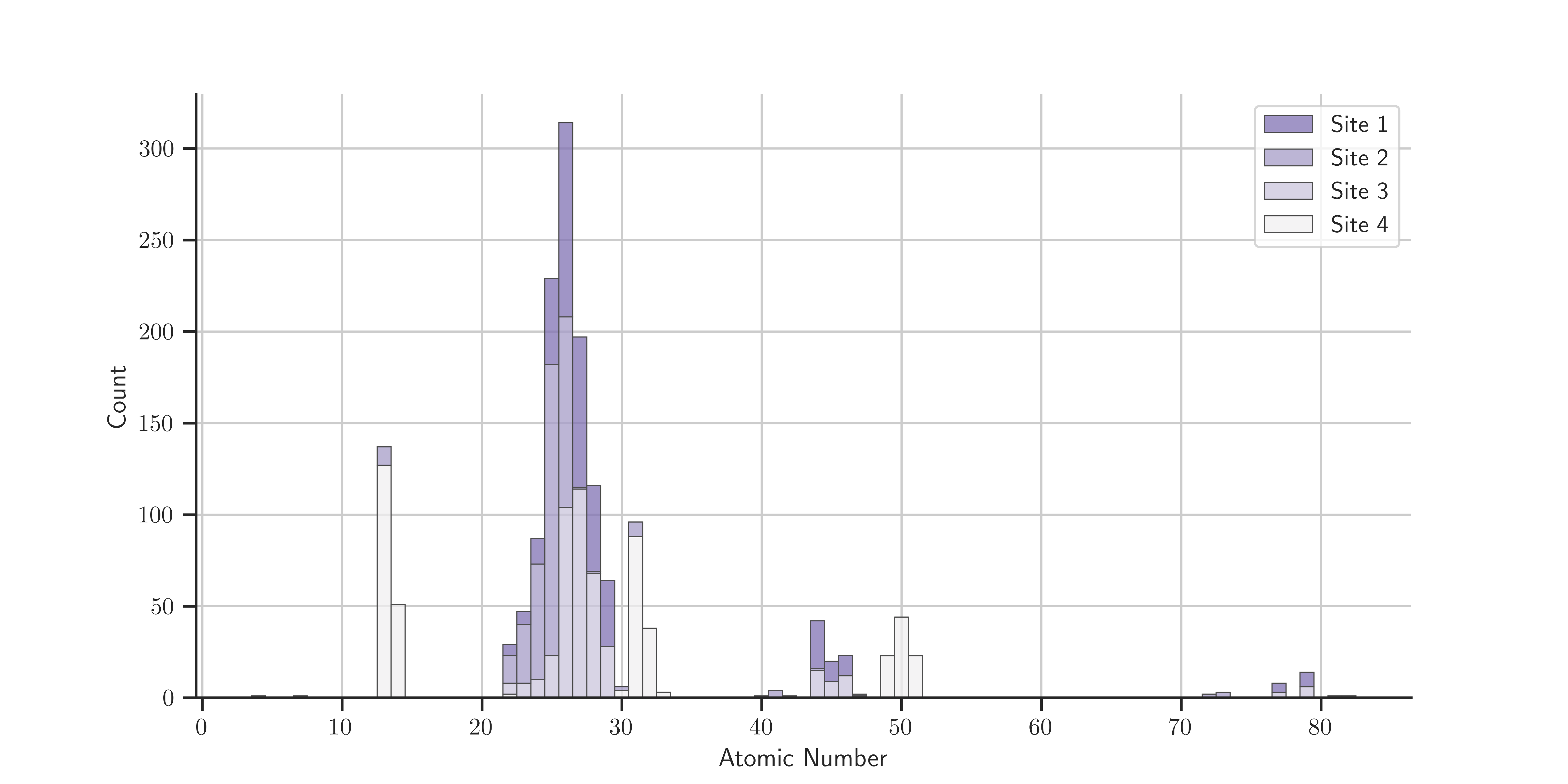}\caption{Distribution of atomic numbers in the GGA data set after processing and cleaning}
\label{Dist}
\end{figure*}

The code of the data 
processing script, as well as the code used to generate the following results and figures, is available~\cite{CodePub}. This allows \textit{e.g.}\ to reevaluate the models if 
more data is added to the JuHemd.
Fig.~\ref{Dist} shows the distribution of atomic numbers across different lattice sites in the Heusler compounds. One can see that Manganese, Chromium, and Iron are contained in a large portion of compounds in the data set. 
\subsection{Model Goals \& Evaluation}
The prediction of $T_\text{c}$ using the descriptors outlined in the previous section leads to a classical regression task. Such regression models aim at predicting $T_\text{c}$ as accurately as possible. 
Different metrics are available to evaluate their performance. The evaluation method of choice is also determined 
by \textit{e.g.}\ the error which is desired to be minimized and the importance and impact of outliers in the prediction. 
The metric used for regression models during this work is the coefficient of determination (Denoted as $R^2$) for test sets, as well as the CV scores. 
$R^2$ measures how well the describing features explain the change in the target variable. Hence, we can be sure to choose a model which properly links the descriptors to $T_\text{c}$. 

Besides the regression, we also transformed our problem into a classification task. 
For the critical temperature, this can be done if one is interested in $T_\text{c}$ to be in a certain range. \textit{E.g.}\ industrial applications~\cite{ApplOfHeus} as magnetic storage devices typically 
require magnetic materials to have a $T_\text{c}$ above $ 60 \degree \mathrm{C}$ in operating conditions at least. To maintain this comfortably and ensure long-time magnetic stability at those 
temperatures, we decided on a threshold of $140\mathrm{K}$ above $60\degree C$ as $T_\text{c}$ for a Heusler compound to be considered as “High” $T_\text{c}$~\cite{400K}. Classification typically represents an easier modeling task, 
as the predictive process is less demanding compared to a regression problem. Hence, if one is only interested in magnetic Heusler alloys, which are candidates for an industrial application, 
but the exact value of $T_\text{c}$ is not of interest in the first place – as the exact value could still be determined in a later step using the established \textit{ab initio} + MC method for the compounds 
classified as potentially relevant – one can stick to classifying model algorithms. This type of classification model can be used to filter a large number of potential compounds to determine 
which should be examined further, \textit{e.g.}\ by a DFT calculation in a high-throughput materials screening context. 

For the classification task, additional considerations on how to evaluate the model performance have to be made. The number of correctly predicted categories would be called the accuracy. However, the errors made in the 
classification do not have the same significance. If a compound is classified as a “low $T_\text{c}$” but truly has a “high $T_\text{c}$” this means the model misses out on a 
material with a potential industrial application. The other error the model can make is classifying a “low $T_\text{c}$” compound as a “high $T_\text{c}$” compound. Which in the worst 
case means a waste of computational resources in the example above. Therefore, the goal for a classification model in this application has to be to minimize data points falsely classified as “low $T_\text{c}$” while still keeping the number of falsely as “High $T_\text{c}$” classified compounds low, in order not to waste too many computational resources on these false positives. 
Hence, we decided to continue with the balanced F1-score, which represents a trade-off between precision and recall.

The model performance is determined using $20\ \%$ of our data as a test data set. This test set has been picked randomized out of the 
whole data set and is used for calculating the test scores only. This gives an insight into how the model would perform on similar but unseen data. 
4-fold CV scores were used in the course of this research in order to perform hyperparameter optimization using a grid search algorithm~\cite{automl,James2013}.
Hence, for this hyperparameter optimization, we again partition the training data into a 20\% validation set for each individual CV fold and use only the remaining 60\% for training.
After the hyperparameter optimization, the validation set is included to train the model using the best-performing hyperparameters before proceeding with the testing.

\begin{figure*}[]
\includegraphics[width=\textwidth]{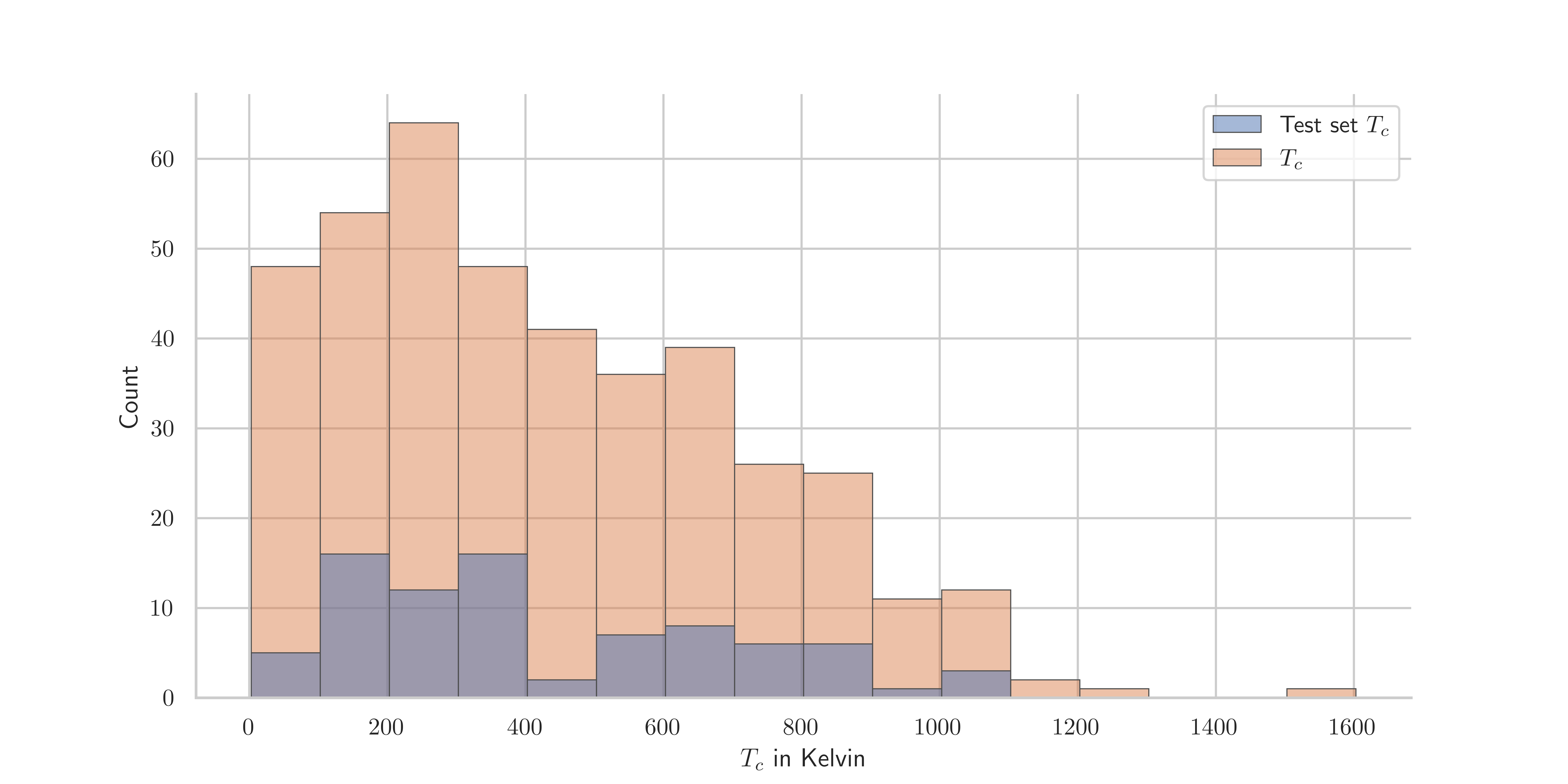}
\caption{Distribution of $T_\text{c}$ values in the total data set as well as in the test set only.}
\label{Test}
\end{figure*}

 The distribution of the $T_\text{c}$ values in the test set is displayed in Fig.~\ref{Test}. The values above 1500 K can be considered as outliers and are hence removed from the data set before the data is used in an ML workflow.

 For all shown scores, it holds: The closer the score is to 1.0, the better the model's predictive performance is. 
\subsection{ML Techniques}

The zoo of ML models and techniques continues to grow year by year. It has already grown to such an extent that it is impossible to cover all possibilities and learning algorithms in a single paper. Hence, we limited our analysis to frequently used and established models. It is also worth mentioning that we excluded neural network models (NNMs) from our research on this data set due to the tabular nature of the data~\cite{NotAllYouNeed,TreeTabData}.

Before training and evaluating models, it is usually not possible to anticipate which model will perform best on a given data set. This is commonly referred to as the "no free lunch theorem"~\cite{Lunch}
Hence, the regression models we evaluated are depicted in the following table:

\vspace{0.25cm}
\resizebox{\columnwidth}{!}{
\begin{tabular}{llll}
\toprule
Linear & Non-linear & Ensemble \\
\midrule
LASSO & K-Nearest Neighbors & Extra Trees \\
LASSOLars & Decision Tree & Random Forest \\
Linear Regression & & \\
\bottomrule
\end{tabular}
}
\vspace{0.25cm}

We have also examined classification models based on similar learning algorithms as some of the regression models depicted in the previous table, as well as a layered indirect classification based on the prediction of the regression models. The indirect classification has been performed to be able to compare the performance of the regression models to their classification counterparts. Since classification is an inherently less complex task than regression, the models would be hard to compare otherwise. The reason underlying this comparison is to determine the best-performing overall model to be used for the feature importance analysis. 
\subsection{Feature Importance}

ML algorithms can be used as black boxes, simply yielding a desired prediction. However, by not applying XAI techniques to understand the model's prediction, we could miss out on the opportunity to improve our understanding of the underlying physics and validate that the model, indeed, has learned physical key properties and relations.
It is considered best practice to perform feature importance analysis using the model which performs best. This is possible by using the SHAP package~\cite{SHAP} including the inbuilt visualization options for the SHAP values. SHAP values represent an ML-specific case of the coalition game theory originated Shapley values~\cite{Shapley}. SHAP values can be considered as the estimated average contribution of an individual feature – given a set of features – to the deviation of a predicted value from the mean prediction. Hence, Shapley values can be interpreted as a “driving force” of individual features away from the mean prediction. This allows us to explain the model's prediction locally for each individual prediction and globally for a set of predictions~\cite{SHAP}. The SHAP package is - in principle - model agnostic but has routines optimized for certain model types such as \textit{e.g.} tree-based model~\cite{SHAPTree}.

\section{Results \& Discussion}
In the following, we showcase the scores and results we achieved in training different ML models. In the spirit outlined in the introduction, we investigated the case in which we used descriptors, including results from the DFT simulations, to only learn the results of the Monte-Carlo step first. In a second, independent analysis, we neglected all descriptors that are only available after the DFT simulation and tried to predict $T_\text{c}$ values by using only the atomic data.

For the classification, we will discuss the best-performing model and differences between direct and indirect classification for both the complete descriptor set as well as the reduced descriptor set.

\label{Res}
\subsection{Complete descriptor set}

\subsubsection{Regression}\label{SecReg}
\begin{figure*}[]
\includegraphics[width=\columnwidth]{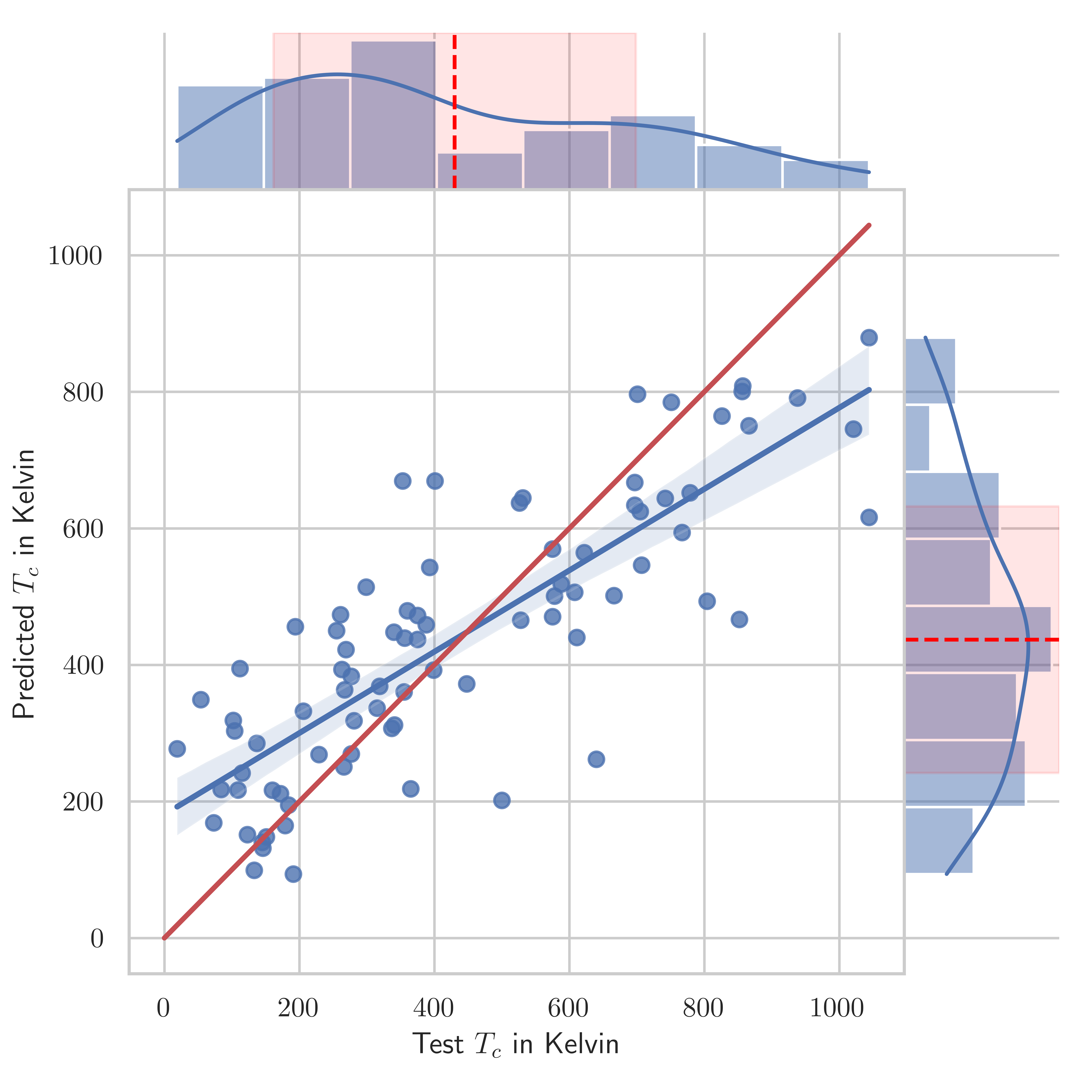}
\includegraphics[width=\columnwidth]{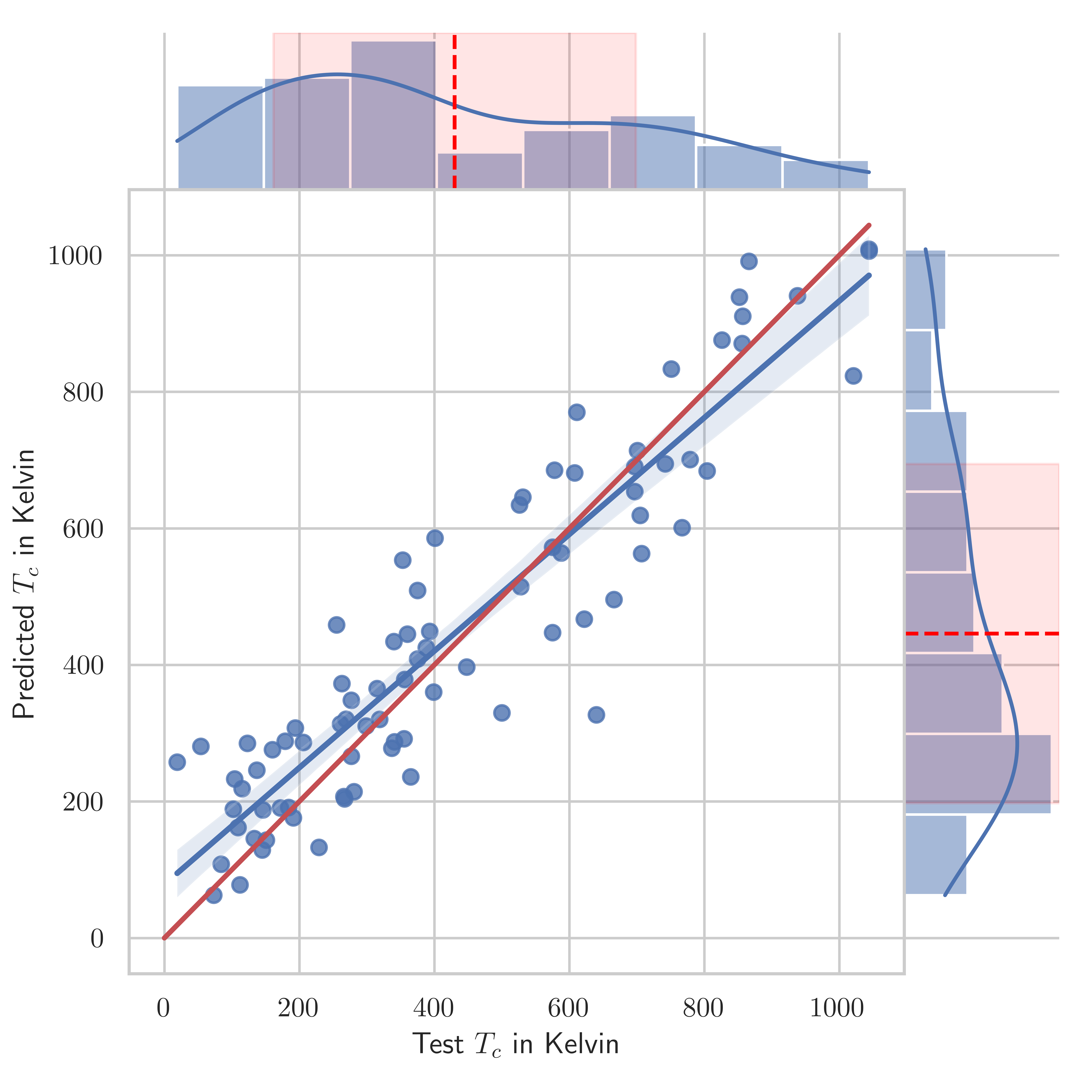}
\caption{Prediction result of two models for the test set. On the left, the data for the linear LASSO model is shown, while the right panel shows the data from the Extremely Randomized Trees (Extra Trees) Regression model~\cite{ET}. The red line indicates a perfect match between the predicted and expected data, the blue line (with shade) a linear regression through the predicted data points (with a 95\% CI envelope for the regression, computed using a bootstrap  \cite{BootstrapCI} based method). On the side distribution plots of the test sets, true $T_\text{c}$ values and the predictions are added.}
\label{ETR}
\end{figure*}

A first impression of the predictive performance of two different regression models can be obtained from Fig.~\ref{ETR}. For a simple linear model (LASSO) as well as a more complex Extremely Randomized Trees (Extra Trees) regression model, we report the predicted value of $T_\text{c}$ in relation to the value obtained from the full simulation for our test set. While the Lasso results show a systematic error by underestimating the higher values of $T_\text{c}$ while overestimating the critical temperature for the low $T_\text{c}$ Heuslers, this deficiency is substantially reduced in the Extra Trees model. In addition, this model also reproduces the distribution of the values much more accurately and shows less spread around the ideal red line.
\begin{figure}[]
\includegraphics[width=\columnwidth]{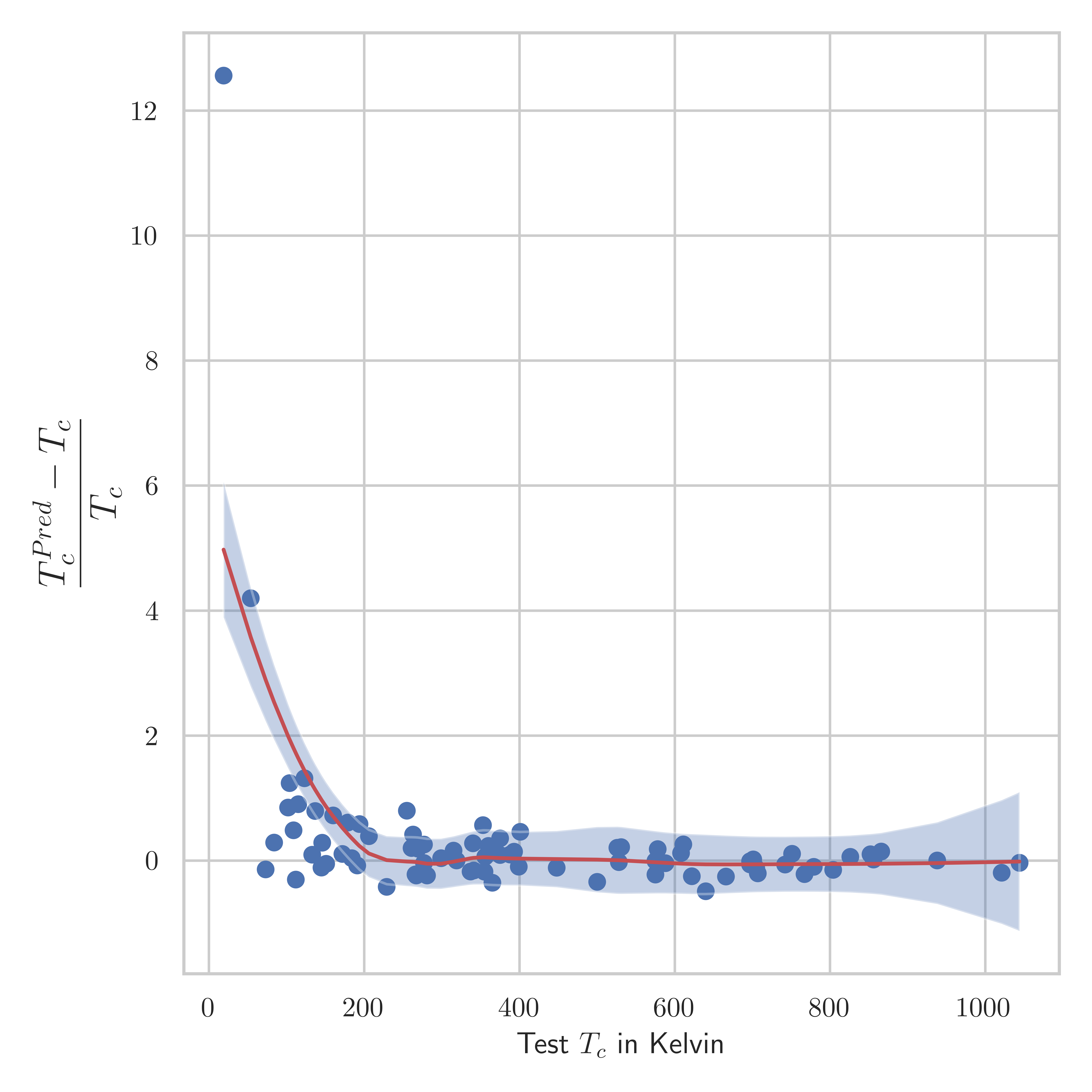}
\caption{Residual depiction of the Extra Trees regression shown on the right of Fig.~\ref{ETR} including a LOWESS smoothing applied to the data points with a pointwise 95 \% CI envelope.}
\label{ETRResid}
\end{figure} 

This can be confirmed by the plot (Fig.~\ref{ETRResid}) of the difference between the simulated (true) value and the prediction, showing a low relative residual error for the whole temperature range except the very low $T_\text{c}$ values. This error for low values arises from the scale of the very low-temperature values, which enlarges the relative error due to its fractional nature.

A more comprehensive overview of the results of various regression models can be obtained from Table~(\ref{LinRegs}), in which we report some performance indicators for the linear, non-linear, and ensemble models.
 \begin{table}
\caption{\label{LinRegs}Regression scores of trained models using the full data set including \textit{ab initio} originated descriptors. The rows show the linear models, the next rows the non-linear models, and the final rows the ensemble models.}
 \begin{ruledtabular}
\begin{tabular}{l l l l l }
 & CV Score & Train $R^2$ & Test $R^2$ \\ \hline
LASSOLars & $<<0$ & 0.77 & 0.65 \\ 
LASSO & 0.66 & 0.78 & 0.66 \\ 
Linear Reg. & $<<0$ & 0.77 & $<<0$ \\ 
Decision Tree Regression & 0.59 & 1.0 & 0.62\\ 
KNN & 0.49 & 0.66 & 0.57 \\ 
Extra Trees & 0.77 & 1.0 & 0.85 \\ 
Random Forest & 0.74 & 0.97 & 0.82 \\ 
\end{tabular}
\end{ruledtabular}
\end{table}

It is clearly visible that the ensemble models are performing best on the test set and in the CV scoring. However, these good predictions are accompanied by a high degree of overfitting to the training data, easily recognizable by the nearly perfect $R^2$ score on the training set, \textit{i.e.}\ a very low or even close to vanishing bias~\cite{James2013}. The fact that only the ensemble models exhibit a reasonably low bias indicates that the complexity of other models does not meet the complexity of the quantity to predict and/or the data. In general, the model complexity has to be adjusted to the data complexity~\cite{DataComp}. It is clear that a simple linear regression, as well as the K-nearest neighbor model, does not meet this requirement in our case. This finding reflects the complexity of the physical processes responsible for the emergence and stability of magnetic phenomena~\cite{Compl}.

 The typical approach to cope with overfitting is increasing the regularization~\cite{Regul}. However, even by applying regularization, we could not determine models with improved CV scores, which itself indicated that a lack of regularization is not the root cause for the overfitting. Moreover, when dealing with different iterations of the data set over the course of this study we observed an improvement of the model performance, \textit{e.g.}\ seen in a decreasing variance, with every increase of the total amount of included Heusler compounds. This indicated that a lack of training data causes a high variance for the more complex models. This also explains the higher test score compared to the CV score. The model in the test case had the full training data available, while for the calculations of the CV score each of the four CV scores – which are depicted here – had only 75 \% of the training set available for training.\\
\subsubsection{Classification}
Since classification is a significantly easier task than a regression, we expect to see an improved model performance for each classification model compared to the regression case on this data set. 
In table \ref{ClassWithDFT}, the results of each linear, non-linear, and ensemble classification model, as well as indirect classification models based on a linear model and an ensemble regression model from section \ref{SecReg}, are displayed.
 \begin{ruledtabular}\begin{table}[]
\caption{\label{ClassWithDFT}Direct and indirect classification scores of a model selection using the full data set, including \textit{ab initio} originated descriptors. The rows show the linear models, the next rows the single-tree-based model and the final rows the ensemble-based model setups.}
\begin{tabular}{l l l l l}
 & CV Score & Train F1 &Test F1  &  Test Accuracy\\ \hline
Logistic Reg. & 0.82 & 0.91 & 0.86
 & 0.89 \\ 
Indirect LASSO & n/a. & 0.86 & 0.81& 0.85 \\ 

Decision Tree& 0.74 & 1.0 & 0.75
 & 0.77 \\ 
Extra Trees& 0.82 & 1.0 & 0.91 & 0.93 \\ 
\multicolumn{1}{l}{\begin{tabular}[c]{@{}l@{}}Indirect Extra \\Trees model\end{tabular}} & n/a. & 1.0 & 0.89 & 0.92
\\ 
\end{tabular}
\end{table} 
\end{ruledtabular}
As expected, the CV scores of the classification models are significantly higher than the scores of the regression models, which corresponds to a lower bias. Similarly, the results for the test set are closer to the ideal prediction, as there is less variance occurring than for the regression models. This aligns with our interpretation of the overfitting in the regression case due to the fact that classification is an easier task, so there is less training data required to fit classification models to the data as the complexity of the quantity to predict is reduced from a continuous quantity to a binary value. This reduction is only possible as we know which minimum $T_\text{c}$ values are required to be relevant to an industrial application. 
\subsubsection{Feature Importance}
\begin{figure}[]
\includegraphics[width=\columnwidth]{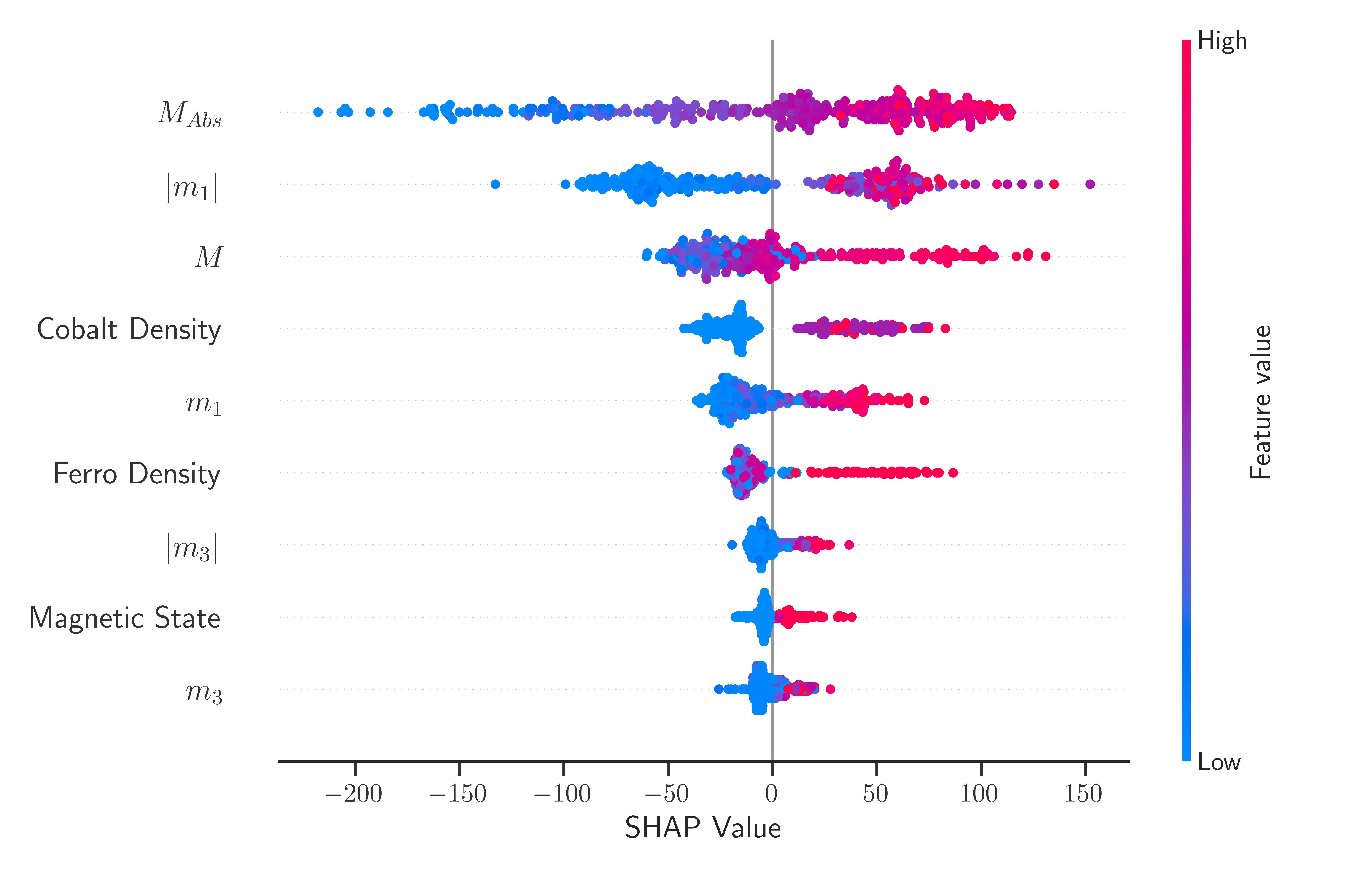}\caption{SHAP beeswarm summary plot of the nine descriptors with the largest SHAP values~\cite{SHAP,Shapley}}
\label{FI}
\end{figure}

After searching for a working set of hyperparameters in section \ref{SecReg} using the training set, we used the determined hyperparameters and chose the Extra Trees Regression~\cite{ET} as our best-performing model to conduct a feature importance analysis using the SHAP package~\cite{SHAP} and the corresponding SHAP values. The SHAP values have been calculated for the training data set. These values for the most relevant features are shown in Fig.~\ref{FI}.
 \\
Besides the SHAP values, the color of the data points encodes the relative scale of the feature for each individual data point. Meaning that if there is a clear horizontal color fade visible, this implies a systematic impact of this feature for the predicted quantity. 

From Fig.~\ref{FI} one can see that for the Extra Trees model, the absolute magnetic moment of the compound has the largest impact on the $T_\text{c}$ prediction. All nine most relevant quantities are either magnetic moments or indirectly related to magnetism (\textit{e.g.}\ the Cobalt density of the compound), which confirms that the magnetic material-specific properties indeed have the largest impact on the value of the critical temperature. While all the nine quantities are positively correlated to $T_\text{c}$, \textit{i.e.}\ have an increasing impact on the $T_\text{c}$ when they increase too. For some of the quantities, this is of course an artifact of our descriptor construction. For example, we encode the magnetic state as an integer, with the smallest possible encoding $000$ denoting that the material forms neither a ferromagnet, an anti-ferromagnet, nor a spin-spiral. In contrast, the fact that the model assigns most significance to the nine quantities listed here was obtained without providing any physical knowledge of the system, besides the fact that we included these descriptors in the first place. Thus, the modeling singled out that these parameters indicate the kind of “physical insight” that can be obtained from ML. For example, the high relevance of the absolute magnetic moment of the compound for the $T_\text{c}$ is of course in line with the relation one would obtain from even very basic physics models of magnetism. 

\begin{figure}
 \centering
 \includegraphics[width=\columnwidth]{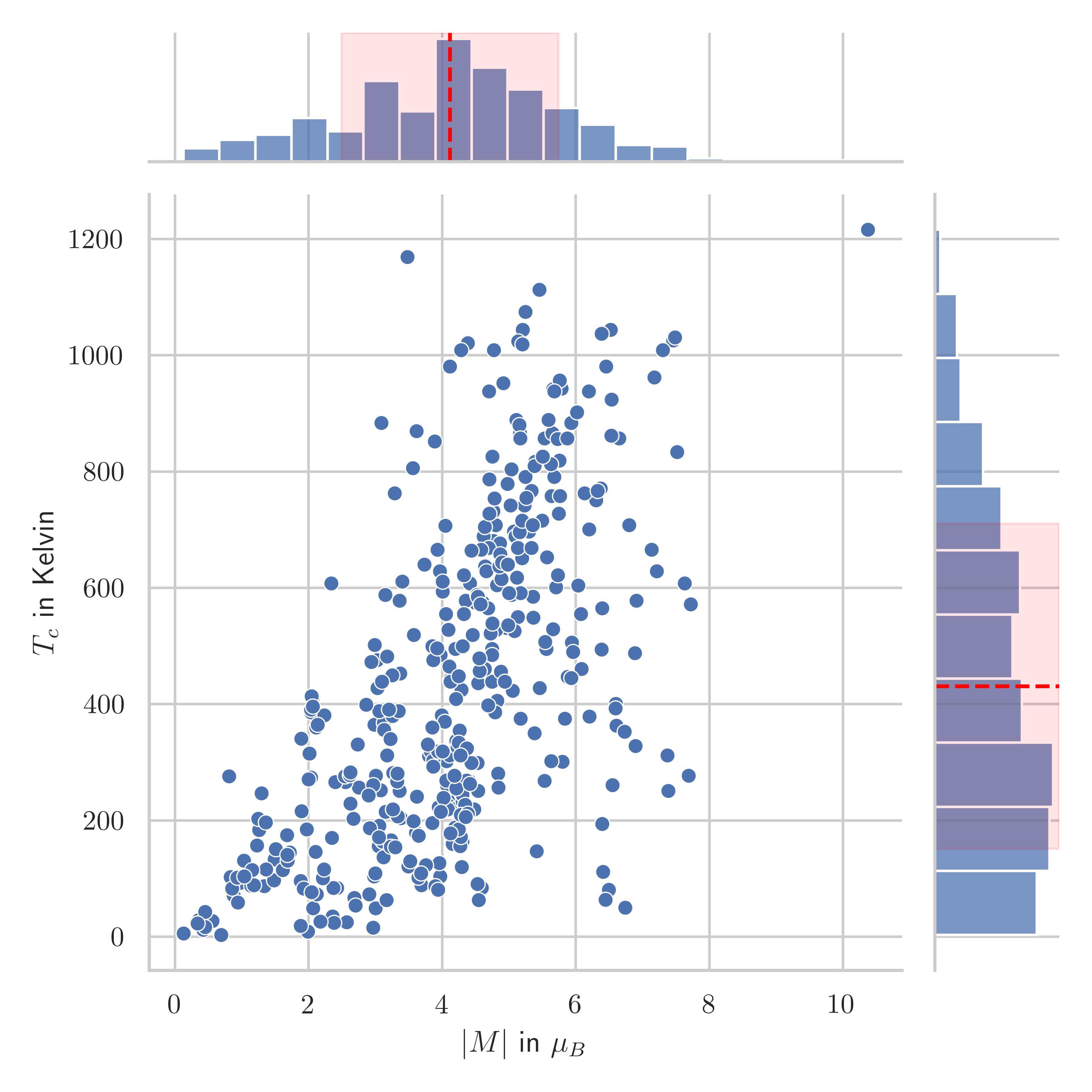}
 \caption{Relation between absolute magnetic moment of the Heusler compounds to their $T_\text{c}$ values for the whole data set.
}
 \label{TcM}
\end{figure} 
Looking at Fig.~\ref{TcM} one can draw even more conclusions. It can be seen that $T_\text{c}$ is not simply proportional to $M_{Abs}$. Instead, Heusler alloys with a higher $M_{Abs}$ can show a higher $T_\text{c}$. However, while a high absolute magnetic moment does not guarantee the emergence of a high $T_\text{c}$, a low $M_{Abs}$ prevents the occurrence of high $T_\text{c}$ values. Therefore, it is safe to say $M_{Abs}$ is acting as an upper boundary: \begin{equation}
    T_\text{c} \leq C M_{Abs}
\end{equation}
\subsection{Data set without DFT-originated descriptors}
\subsubsection{Regression}
Retraining the Extra Trees Regression model as well as the LASSO model to the reduced descriptor set and again performing hyperparameter optimization using a grid search algorithm, we achieved the regression scores displayed in table \ref{RegsNoDFT}. 
 \begin{ruledtabular}
\begin{table}[]\caption{\label{RegsNoDFT}Regression scores of promising models using the reduced data set, excluding \textit{ab initio} originated descriptors.}
\begin{tabular}{l l l l}

 & CV Score& Train $R^2$& Test $R^2$ \\ \hline
Extra Trees & 0.52 & 1.0 & 0.76 \\ 
LASSO & 0.31 & 0.58 & 0.63 \\ 
\end{tabular}
\end{table}
 \end{ruledtabular}

As expected, one can observe a clear decrease in performance compared to the case where DFT-originated descriptors have been used. In particular, the LASSO results now have huge deviations such that one could question its fitness for any practical application. Therefore, we can already conclude, that a prediction of the critical temperature without the use of the basic magnetic properties predicted by a DFT simulation is not really possible in our scenario. Therefore, we decided not to analyze this further, but to concentrate on the easier classification task.
\subsubsection{Classification}\label{SecClassNODft}
The achieved classification model results using no DFT-originated descriptors at all are displayed in table \ref{ClassNoDFT}. This table contains exactly the same models as seen before in table \ref{ClassWithDFT}.
 \begin{ruledtabular}
\begin{table}[]\caption{\label{ClassNoDFT}Direct and indirect classification scores of a model selection using the reduced data set, excluding \textit{ab initio} originated descriptors. The rows show the linear models, the next rows are the single-tree-based model, and the final rows are the ensemble-based model setups.}
\begin{tabular}{l l l l l }
 &CV Score & Train F1 & Test F1 &  Test Accuracy\\ \hline
Logistic Reg. & 0.68 & 0.75 & 0.75
 & 0.79 \\ 
Indirect LASSO & n/a. & 0.75 & 0.8
 & 0.88 \\ 
Decision Tree& 0.66 & 1.0 & 0.8
 & 0.84 \\ 
Extra Trees & 0.74 & 1.0 & 0.84 & 0.87 \\ 
\multicolumn{1}{l}{\begin{tabular}[c]{@{}l@{}}Indirect Extra \\Trees model\end{tabular}}  & n/a. & 1.0 & 0.91 & 0.93
\\ 
\end{tabular}
\end{table} \end{ruledtabular}
From the test set of 82 compounds, our constructed indirect Extra Trees classification model managed to correctly classify 47 “Low” $T_\text{c}$ and 29 “High” $T_\text{c}$ compounds. Of each class, 3 compounds have been wrongly predicted. We consider falsely classifying a “Low” $T_\text{c}$ compound as a “High” $T_\text{c}$ not so relevant for practical application. The worst outcome in a potential use case is that the model suggests a “High” $T_\text{c}$ compound, and when computing it using a more sophisticated – and hence computationally more intensive – approach, one finds that the “High” $T_\text{c}$ label has been falsely assigned. However, if a “High” $T_\text{c}$ compound is classified as “Low” $T_\text{c}$ in a high-throughput screening process it will probably never be computed with a more sophisticated approach, which causes this compound to be potentially “lost” for future research. In the case of this model, we saw that this crucial error for falsely classifying a “High” $T_\text{c}$ Heusler as a “Low” $T_\text{c}$ Heusler is below 5\% and hence meets typical confidence criteria. This concludes that the indirect Extra Trees classification is capable of classifying the $T_\text{c}$ in “High” and “Low” values even without the DFT-originated data. While “Low” means $T_\text{c}$ is too low to be relevant for industry applications. 
\subsubsection{Feature Importance}
\begin{figure}[]
\includegraphics[width=\columnwidth]{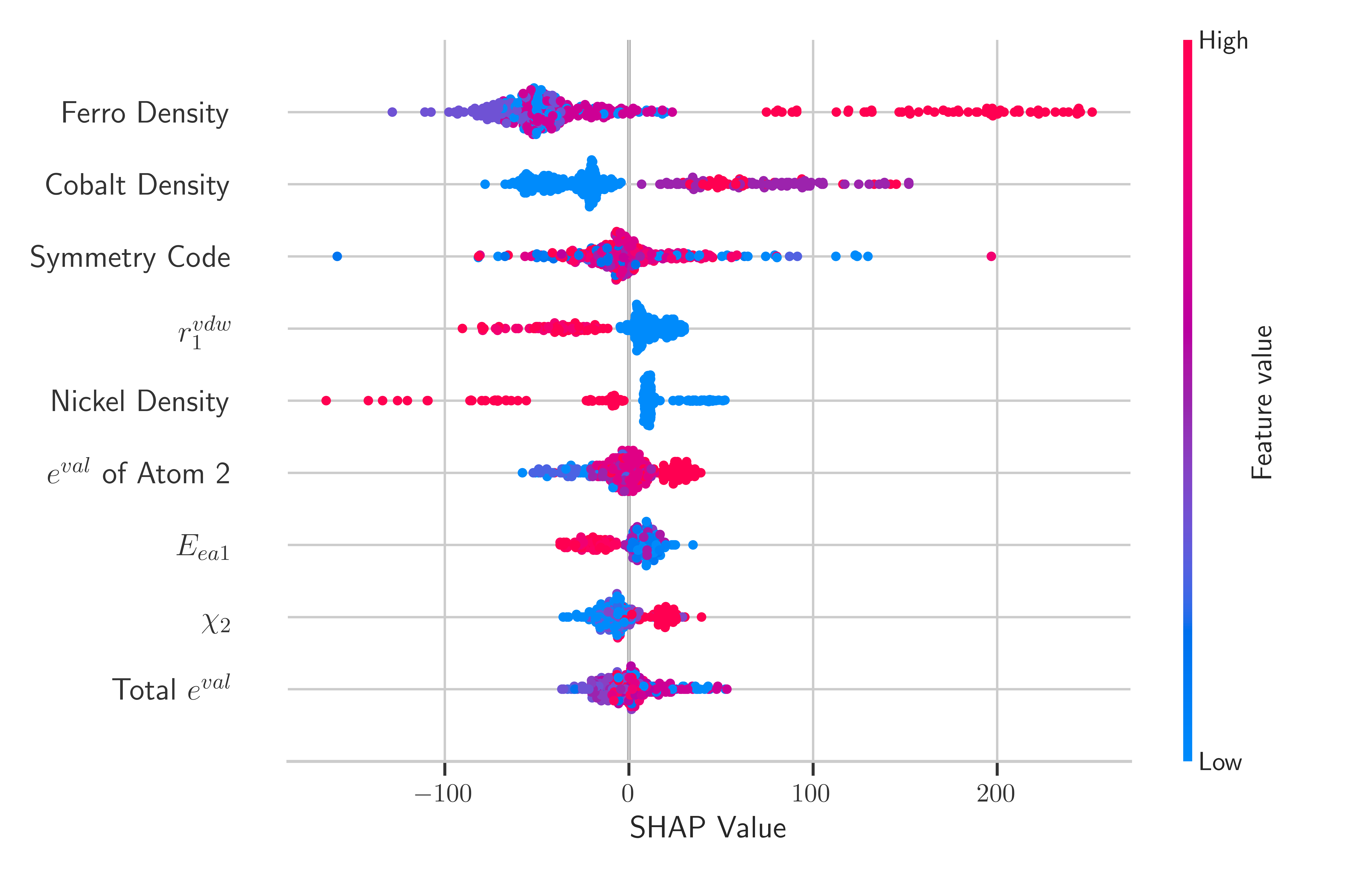}\caption{SHAP beeswarm summary plot of the nine descriptors of the data set without the DFT-originated data with the largest SHAP values.}
\label{FINODFT}
\end{figure}

Computing the SHAP values for the reduced descriptor set and visualizing them as we did before results in the beeswarm plot shown in Fig.~\ref{FINODFT}.
As we can see in Fig.~\ref{FINODFT}, removing the DFT-originated descriptors and, therefore, the calculated magnetic moments caused other quantities to become more impactful. As one can expect, these are very closely related to the magnetic moments (\textit{e.g.}\ the density of ferromagnetic materials in the compound as well as the Cobalt and Nickel densities). However, now we observe more complex relations than in the previous feature importance plot, demonstrating the lower significance of these quantities for the critical temperatures. We can see a negative correlation between the van der Waals radius of the atom on site one ($r^{vdw}_1$), the Nickel density in the compound, and the electron affinity of the atom on site one ($E_{ea1}$) with a decreasing $T_\text{c}$ as these quantities increase. For the fraction of ferromagnetic atoms, the effect is much less obvious. We can see that very high densities of ferromagnetic atoms in the compound contribute to a largely increased $T_\text{c}$ prediction. However, on the other hand, a low density of ferromagnetic atoms does not lead to an equally decreased prediction of $T_\text{c}$. Interestingly, this reflects our previous result that a large absolute magnetic moment corresponds to an upper boundary for the $T_\text{c}$. Since a large amount of ferromagnetic compound constituents is highly correlated with a large magnetic moment. The required and obtained model complexity is also observed in the SHAP values of the symmetry code. Since this is an arbitrary-ordered label for the symmetry group of the compound, there is no clear order of the feature value that correlates with the $T_\text{c}$. However, the model seems to have learned that some feature values have a larger impact on $T_\text{c}$ than others, which is indeed possible.

\begin{figure*}[htp]
\centering
\includegraphics[width=.33\textwidth]{"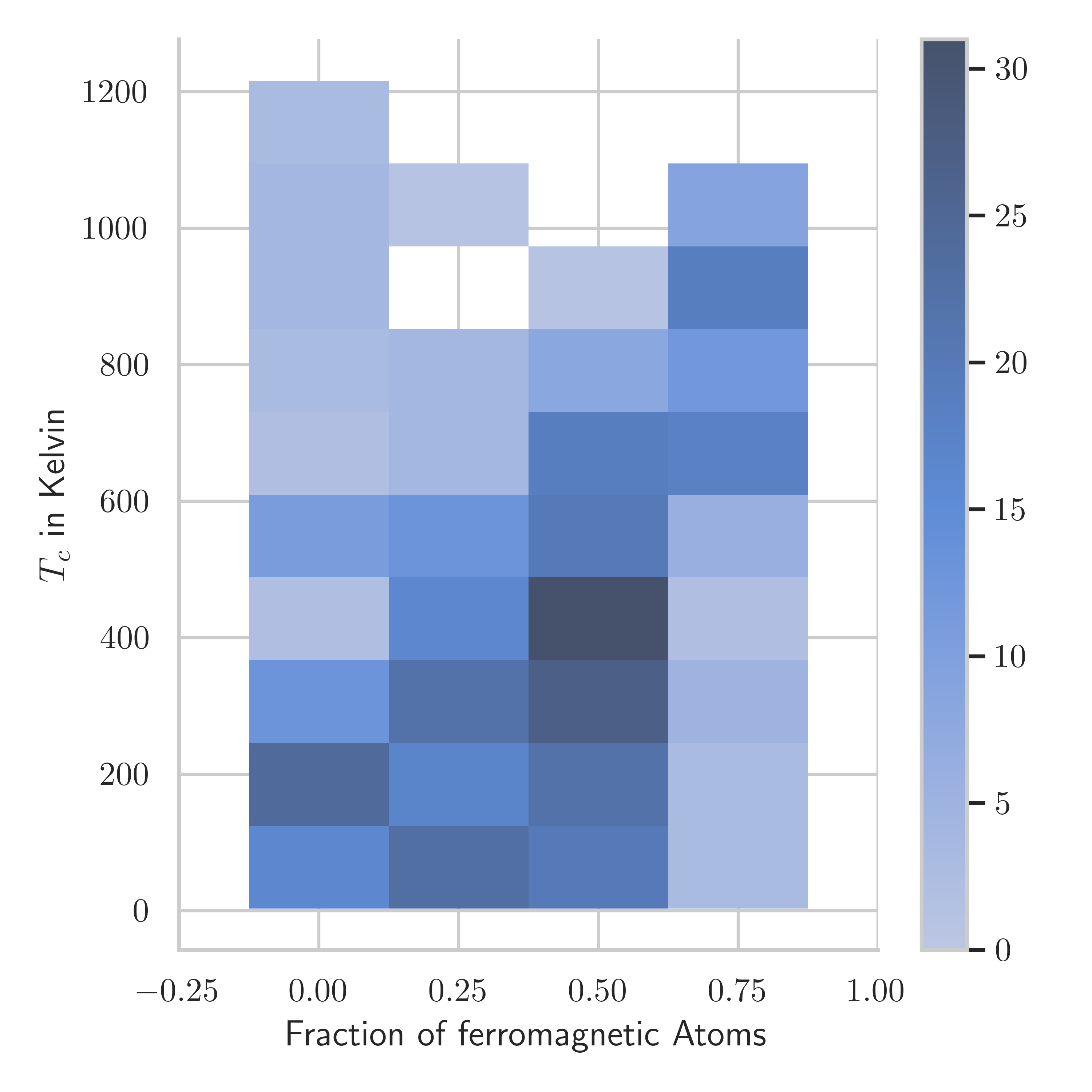"}\hfill
\includegraphics[width=.33\textwidth]{"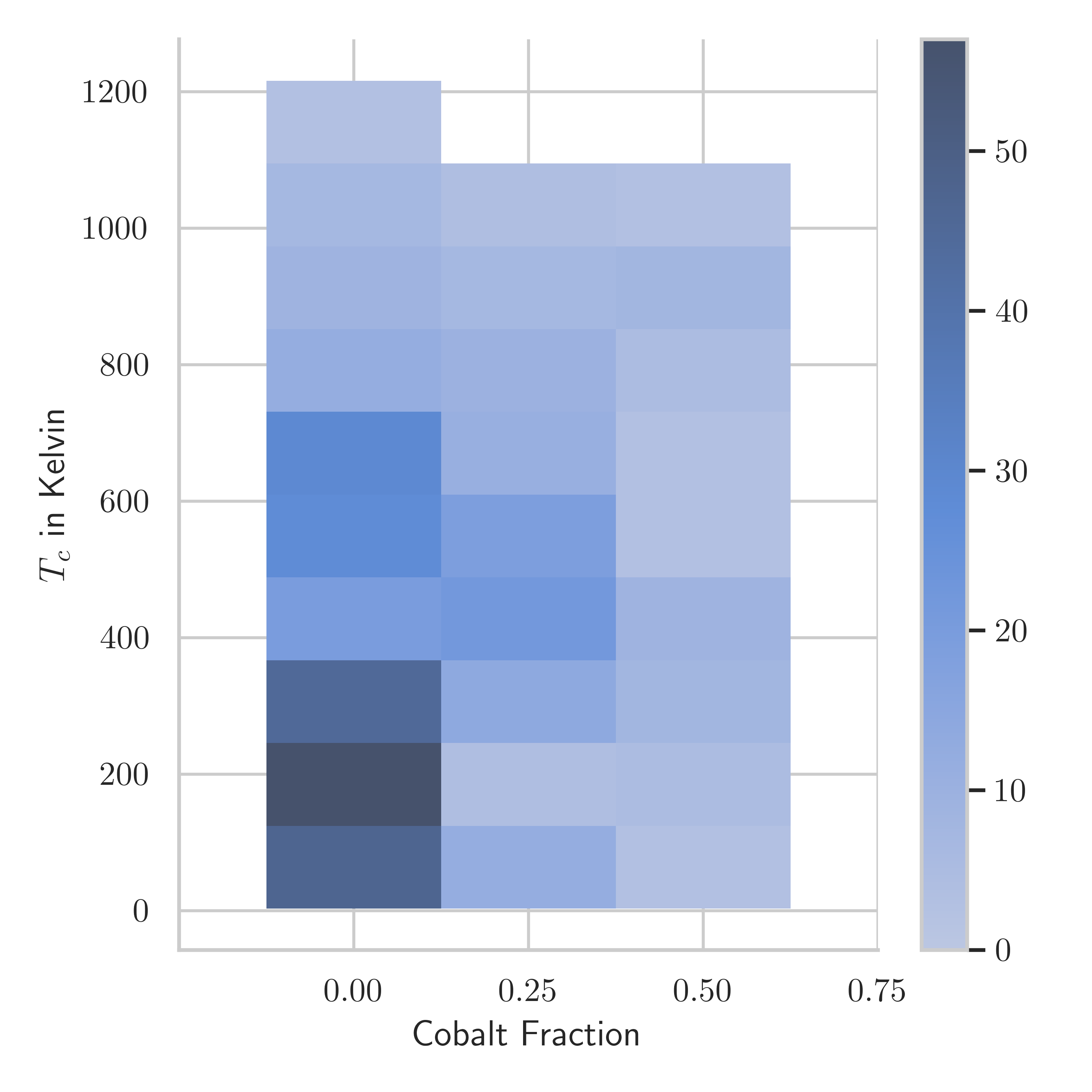"}\hfill
\includegraphics[width=.33\textwidth]{"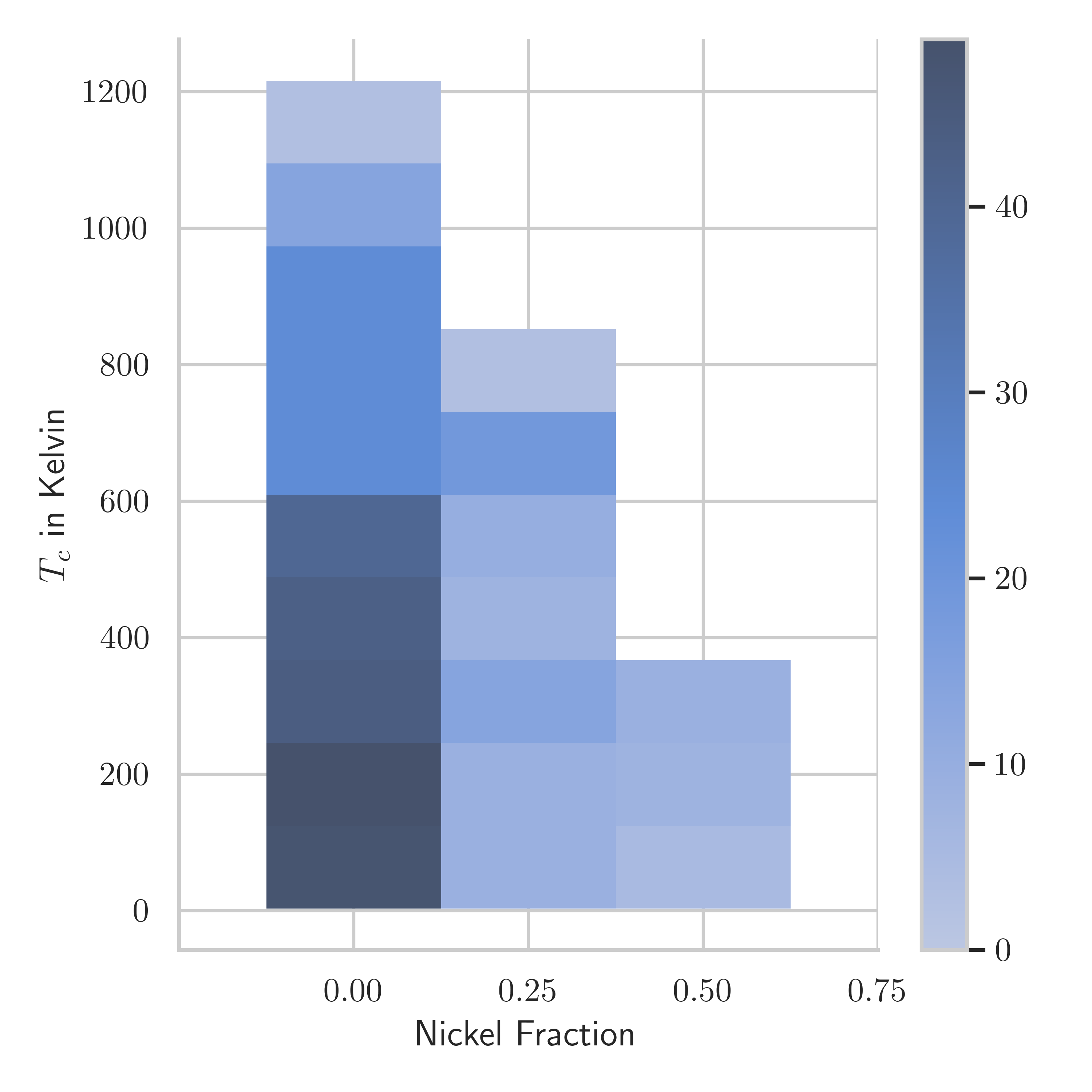"}
\caption{Relation between $T_\text{c}$ and the fraction of ferromagnetic elements (left), cobalt atoms (center) and nickel (right) shown as a heatmap style histogram. The darker the color, the more compounds can be found in the colored region.}
\label{DensitiesE}
\end{figure*}

As the density of the ferromagnetic atoms, the cobalt and nickel atoms turn out to be relevant quantities in Fig.~\ref{FINODFT} we investigated their correlation with $T_\text{c}$ in more detail as depicted in Fig.~\ref{DensitiesE}.
The depicted fractional density histograms confirm the trends we were hinted at by the SHAP beeswarm plot. It is easily visible that a large density of ferromagnetic atoms in the compound is indeed contributing to a larger $T_\text{c}$ value, with one exception: The anti-ferromagnetic case. We can see that there are a few materials that have no ferromagnetic atoms in the compound at all but still a very high $T_\text{c}$. These are strong anti-ferromagnets. This finding can be related to our previous result for the modeling, including the DFT-based magnetization values, in which we have seen that the SHAP values of $M_{Abs}$ hint at a larger impact than those of $M$. As the anti-ferromagnetic compounds have a vanishing $M$ but a large $M_{Abs}$ while resulting in a stable anti-ferromagnetic state with a large $T_\text{c}$. The same relation is, in principle, true for the cobalt density. However, there are fewer compounds containing cobalt in the data than iron. The Nickel density has – for increasing densities – a negative impact on $T_\text{c}$ according to Fig.~\ref{FINODFT}, and as we can see in Fig.~\ref{DensitiesE}, this also emerges from the data. It seems that the presence of Cobalt is not as helpful in contributing to a stable magnetic state as \textit{e.g.}\ Iron.

\section{Summary and Outlook}
\label{Sum}

This work can be seen as a small-scale sandbox-type case study in which lightweight ML algorithms can add value to existing \textit{ab initio} data and eventually replace costly computational steps in layered calculation workflows in the future.

It was demonstrated that qualitative predictions for material-specific properties are achievable with very small errors, even for the limited data set sizes common in materials science. Also, the expectation that the quantitative prediction is much more difficult and requires descriptors with much higher predictive power, has been confirmed. However, we could also demonstrate that even very simple and readily available descriptors not based on any actual calculation in combination with sufficiently complex models could be utilized in a classification task typically part of any high throughput screening.
As demonstrated in this paper, there is a potential use for ML methods in materials science, even in quantitatively predicting properties as complex as the $T_\text{c}$. It is imaginable to perform similar studies on existing data sets of other material families beyond the Heusler alloys. However, one has to consider that the structural homogeneity of the material class we studied here simplified the complexity of the modeling task. This implies that if one would translate the methodical insights gained from this data set of Heusler alloys to a different material class, that there should either be a similar structural homogeneity or if the structural complexity is increased one also has to choose descriptors with equally elevated descriptive value.

By performing feature importance analysis with XAI techniques – such as SHAP values – we gained physical insights about the relations of the target quantity to the included features, as well as the determining properties of the studied material class given in the examined data set. Such analysis can provide a link between a complex ML process with a hard-to-expose underlying mechanism and true physical insight and the gain of knowledge of the system. In this study, be rediscovered dependencies expected from simple physical models without actually providing such knowledge to the process.

Finally, we would like to stress that the methodical approach described in this paper is not limited to predicting $T_\text{c}$ or any other magnetic quantity, but that it can be transferred to any other material-specific property. 
 We believe it is even possible to discover that known materials have currently unknown properties using predictive modeling. 
\begin{acknowledgments}
This work was performed as part of the Helmholtz School for Data Science in Life, Earth and Energy (HDS-LEE) and received funding from the Helmholtz Association of German Research Centres.
 Since parts of the data processing has been performed and the displayed visualizations have been created using dedicated open-source packages, we acknowledge them here~\cite{numpy,mendeleev2014,Tikz,MPL,MPLV,Seaborn,Loess,scikit-learn,stats}. 

We acknowledge Stefano Sanvito for the inspiration to represent the fractional density of each atomic number in the data set as a standalone descriptor, which we gathered from one of his talks. 

The authors thank Fabian Lux for initially hinting us at the Extra Trees regression model. We acknowledge Dirk Witthaut for pointing out the advantages of SHAP values in XAI in comparison to model bound feature importance methods. 
The authors thank Roman Kováčik for discussing the structure and contents of the data published in Ref.~\onlinecite{DataHeusler2022}. Hence, we acknowledge the computing time granted by the RWTH Aachen University (Project: jara0182) which was required in order to collect the data. As the original database as well as the ML-ready data was hosted by Materials Cloud, we acknowledge them \cite{MaterialsCloud}. 

\end{acknowledgments}

\bibliography{Paper.bib}

\end{document}